# Role of substrate on interaction of water molecules with graphene oxide and reduced graphene oxide

*Roman Strzelczyk[1,2], Cristina E. Giusca[1*], Francesco Perrozzi[3], Giulia Fioravanti[3], Luca Ottaviano[3], Olga Kazakova[1]*

[1] National Physical Laboratory, Hampton Road, Teddington TW11 0LW, United Kingdom

[2] Institute of Molecular Physics, Polish Academy of Sciences, Mariana Smoluchowskiego 17, 60-179 Poznań, Poland

[3] Dipartimento di Scienze Fisiche e Chimiche, Universita` dell'Aquila & CNR-SPIN (UOS L'Aquila), L'Aquila 67100, Italy

## Abstract

We study local electronic properties of graphene oxide (GO) and reduced graphene oxide (RGO) on metallic (Pt) and insulating ($Si_3N_4$) substrates in controlled humidity environment. We demonstrate that the supporting substrate plays a crucial role in interaction of these materials with water, with Pt making both GO and RGO insensitive to humidity variations and change in environment. On the other hand, in the case of $Si_3N_4$ substrate a significant difference between GO and RGO with respect to humidity variations is demonstrated, indicating complete water coverage at ~60% R.H for RGO and ~30% R.H. for GO. Irrespective of the substrate, both GO and RGO demonstrate relative independence of their electronic properties on the material thickness, with similar trends observed for 1, 2 and 3 layers when subject to humidity variations. This indicates a relatively minor role of material thickness in GO-based humidity sensors.

## 1. Introduction

Graphene oxide (GO) is a disordered form of graphene containing functional oxygen groups, such as hydroxyl and epoxy, covalently attached to the basal $sp^2$ plane of graphene, as well as carboxylic groups at the edges [1, 2, 3, 4, 5]. Presence of these groups turns the material into an electrical insulator. Reduction of the oxygen-containing groups by a relatively simple thermal or chemical treatment results in reduced GO (RGO) [5]. This procedure allows to dramatically change the electrical properties of the material, converting it from an insulator to a semi-metal [1, 4, 5].

---

[*] Corresponding author. E-mail: cristina.giusca@npl.co.uk

Extensive research interest in GO and RGO owns to relatively easy and cheap production of the material, which is highly attractive for practical applications. For example, superpermeability of GO membranes to water molecules [6] opens a vast field for applications in water remediation and filtration [7]. In particular, GO functionalised by magnetite nanoparticles allows for magnetic-field assisted filtration of polluted water from dangerous substances containing heavy metal cations, see e.g. [8,9].

The large number of the reactive functional groups at the surface makes GO inherently attractive for environmental and biomedical sensing applications where such groups act as active anchoring sites for molecules [19, 20, 21, 22, 23]. On the other hand, the dielectric nature of GO is an obvious disadvantage for a sensor when used as a part of the electrical circuit [5]. As a compromise between graphene and GO in terms of chemical reactivity and electrical properties, RGO provides both the presence of reactive sites necessary for chemical reactivity [10, 11, 12] as well as relatively high electrical conductivity. Additional advantages of GO/RGO for gas sensing include the material's inherent flexibility, transparency and suitability for large-scale manufacturing. In particular, the specific response of GO to $NO_2$ was studied in various conditions and the sensitivity threshold of ~20-50 ppb was demonstrated [see, e.g. Refs. 5, 13, 14]. The sensing ability was attributed to the presence of oxygen functional groups. Additionally, exceptionally high sensitivity to $NH_3$ (~1 ppb), accompanied by the large resistance change and fast sensor response, and combined with a selective response to many other chemical compounds [15] was reported in GO reduced by pyrrole.

Considering even more practical applications, bendable and washable electronic textile gas sensors made of yarn coated by RGO were recently demonstrated [16]. Such sensors demonstrate excellent chemical durability and mechanical stability combined with a high response to $NO_2$ and selectivity to acetone, ethanol, ethylene, and $CO_2$. Other advantages of e-textiles include pliability, a light weight and low cost [17], which makes them desirable materials for high-performance flexible gas sensors [18]. GO was also successfully used in biosensing applications, e.g. protein [19], DNA [20] and enzymatic [21] detection, for regulation of neuronal signalling without affecting cell viability [22], as a biometric live-cell sensor [23] and many others.

A generally accepted electrical mechanism for gas sensing is consistent with the p-type conductivity, where a positively charged vacancy is created in the carbon lattice due to a negative charge transfer (a fraction of electron charge) from GO functional groups to the molecule. Additionally, unperturbed patches of graphene lattice can contribute through the intrinsic quantum capacitance of graphene.

In the case of water vapour, the dipolar nature of water molecules also adds to the overall spectrum of electrical effects [5, 24, 25]. The interaction of GO with water is particularly important due to the observed superpermeability of GO membranes [6] to water molecules as well as restricting conditions, which the presence of water imposes on gas sensing. In a number of cases, poorer sensitivity to gases is observed in humid conditions compared to dry gases, implying that water molecules form dipoles on the GO surface. The sensing mechanism in this case is based on the fact that reaction of water molecules with oxygen-containing functional groups releases protons, leading to an increase of conductivity [12, 26]. Ultrafast GO-based humidity sensors were demonstrated, exhibiting such features as transparency, flexibility and easy upscaling, which makes them extremely attractive for applications [27, 28]. In particular, fast, touch-free sensors responding to modulation of moisture in a user's breath were demonstrated by Nokia for recognition applications [29].

Recently functional scanning probe microscopy was successfully used for studies of local electrical properties of various graphenes and 2D materials [30, 31, 32, 33, 34, 35], e.g. GO/RGO [36, 37, 38] These methods allow to study variations of properties on the nanometer scale, i.e. addressing differences in number of layers, non-homogeneous carrier density, screening and edge effects, local defect structure, etc. Knowledge of local electronic properties (e.g. obtained by scanning Kelvin probe microscopy (SKPM)) provides crucial information about the device performance, e.g. intrinsic resistivity, contact resistance, etc. [39]. Calibration of surface potential measurements and accurate determination of local work function were demonstrated [32, 35]. In GO, correlation of work function and oxygen content [40] as well as their evolution under UV-radiation related to degradation of oxygen groups [41] were studied by SKPM.

In some cases, authors paid appropriate attention to hydrophilicity of GO and to the fact that water molecules form dipoles on the GO surface, thus affecting the correct determination of the local electrical properties (i.e. surface potential or dielectrical constant) [37,42]. Nevertheless, scanning experiments performed in a well-controlled environment and on well-defined samples are rare. In many previous studies combining local electrical measurements with a specific gas environment or level of humidity [43, 44] crucial factors such as film thickness, level of non-specific contamination and pre-conditioning of films helping to get rid of physisorbed molecules (e.g. by vacuum treatment and/or thermal annealing) were not considered, potentially leading to erroneous conclusions and generalization. Additionally, the effects of underlying substrates and related charge transfer were rarely taken into account. Furthermore, previous measurements of GO work function gave a very broad spread of the results, partly because work function is poorly defined in an insulator, but mainly due to unoptimised experimental conditions, i.e. lack of proper grounding of the sample, leading to charging of GO flakes.

The purpose of the current study is to investigate the effect of variable humidity (0–70 % RH) on the local electronic properties (i.e. surface potential variation) of one-, two- and tri-layer GO and RGO in a controllable manner, by means of SKPM. Additionally, we explore the role of supporting substrates (insulating vs metallic) in the charge transfer, as well as the cumulative effect of the substrate and environmental doping on the properties of GO/RGO. Understanding the role of water and taking into consideration GO thickness, will significantly facilitate development of various environmental and biological sensors.

## 2. Experimental

### 2.1 *GO Synthesis and reduction*

GO sheets were prepared via a modified Hummers method [45] and deposited by spin coating on 30 µm wide platinum mesa structures, with 30 µm lateral spacing in between electrodes, prefabricated on top of $Si_3N_4$ substrates. Reduction of GO sheets was achieved by thermal annealing of the samples in ultra-high vacuum at 300 °C. Typical lateral dimensions of GO and RGO sheets obtained via this method range from 40 to 100 µm, thus ensuring that a large number of individual flakes were placed both on $Si_3N_4$ and Pt sections of the substrate, as confirmed by SEM, allowing to study their properties on both metallic and insulating substrates. Metallic mesa structure also allows for efficient grounding of the GO and RGO flakes, which is crucial, in particular, for conducting SPM measurements in vacuum.

### 2.2 *Scanning Electron Microscopy*

SEM images were obtained using the InLens detector of a Zeiss FESEM Supra system operated at 2 kV.

### 2.3 *Raman spectroscopy*

Raman intensity maps were obtained using a Horiba Jobin-Yvon HR800 system and a 532-nm laser excitation wavelength with 1 mW incident power. Data were taken with a spectral resolution of $(3.1 \pm 0.4)$ cm$^{-1}$ and lateral resolution of $(0.4 \pm 0.1)$ µm.

*2.4 SKPM*

SKPM experiments were conducted in an environmental chamber using an NT-MDT Ntegra Aura SPM system and highly doped Si probes (PFQNE-AL, Bruker). Single-pass frequency-modulated SKPM (FM-SKPM) technique was used in all measurements. FM-SKPM operates by detecting the force gradient (*dF/dz*), which results in changes to the resonance frequency of the cantilever. In this technique, an AC voltage with a lower frequency ($f_{mod} \approx 3$ kHz) than the resonant frequency of the cantilever ($f_0 \approx 300$ kHz) was applied to the probe, inducing a frequency shift of $f_0 \pm f_{mod}$. The PID feedback loop monitors the side modes, $f_0 \pm f_{mod}$, and compensates them by applying an offset DC voltage. By measuring the DC voltage at each pixel (equivalent to the contact potential difference ($U_{CPD}$)), a surface potential map is constructed. The same scanning probe (PFQNE-AL) was used for the entire sequence of the measurements presented in the paper.

*2.5 Environmental conditions*

FM-SKPM measurements in variable environmental conditions were performed in a controlled SPM chamber by monitoring the surface potential changes in the following sequence: (i) ambient (temperature and humidity of the laboratory ambient air: T = 21 °C and 40% RH, respectively); (ii) vacuum (P ≈ 1 × 10$^{-5}$ mbar); (iii) vacuum annealed state (with the sample cooled down to room temperature following thermal annealing at 150 °C for 1 hour); (iv) dry $N_2$ atmosphere (research grade 99.9995% purity); (v) water vapour balanced with $N_2$ with relative humidity in the range 10 -70%. For humidity exposure, RH was increased in a stepwise manner from 10% to 70% and the surface potential measurements were carried out immediately after reaching the target humidity level (≈5 min at low RH and up to ≈30 min for 70% RH).

**3. Results and discussion**

The morphology and size distribution of as-prepared and reduced GO samples are first studied using SEM (Figure 1), which confirms that a large number of both GO and RGO sheets rest both on the $Si_3N_4$ substrate and the Pt electrodes. Individual GO and RGO flakes in varying environmental conditions have been further studied using AFM/SKPM as described in the next sections. Raman spectroscopy was used to assess the degree of reduction of GO flakes and results are presented in the Supplementary Information (S.I.) section.

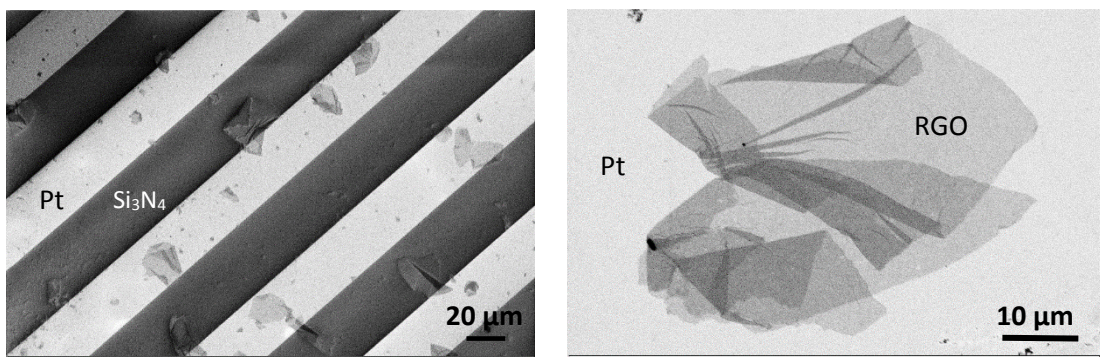

Figure 1: a) Representative SEM image of GO flakes on Si$_3$N$_4$ substrate with Pt electrodes. Pt electrodes show bright contrast in the image; b) Individual RGO flake on Pt substrate.

3.1 *Reduced graphene oxide – humidity variations*

Figure 2a shows a topographic AFM image of RGO flakes, overlaying the border step (denoted by a dashed line) between silicon nitride and platinum areas of the substrate. The area of interest is chosen such that domains of different thickness (i.e. 1-3 layer (L) RGO) are positioned both on Si$_3$N$_4$ and Pt areas, as indicated in the figure. Figures 2b-f show representative surface potential maps of the same area obtained in different environmental conditions, i.e. from ambient to vacuum, to the vacuum annealed state and then from low (10% R.H.) to high (70% R.H.) relative humidity. The humidity level was gradually increased in successive 10% RH steps. All measurements are taken at room temperature.

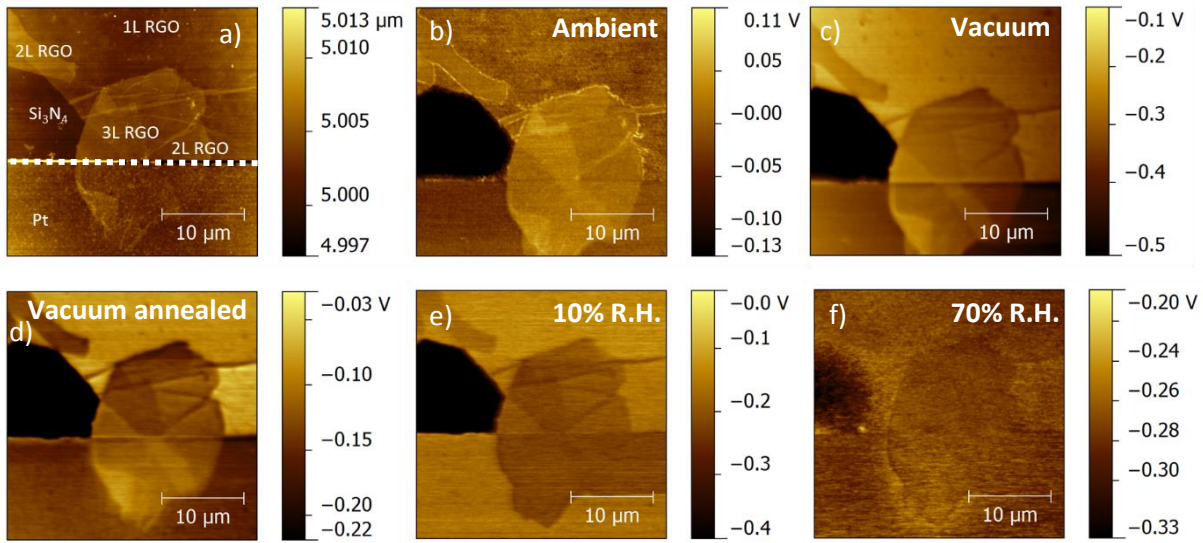

Figure 2: a) Topography and b-f) surface potential maps of RGO at different humidity levels: b) ambient; c) vacuum at RT; d) after thermal annealing in vacuum; e) 10% R.H. and f) 70% R.H. Underlying substrate and number of RGO layers are shown in a). Dashed line in a) indicates the border between $Si_3N_4$ and Pt areas of the supporting substrate.

Values of the relative surface potential (i.e. the surface potential difference between individual domains and either $Si_3N_4$ or Pt substrates) extracted from the images are summarised in Figures 3a and b with reference to $Si_3N_4$ and Pt substrates, respectively.

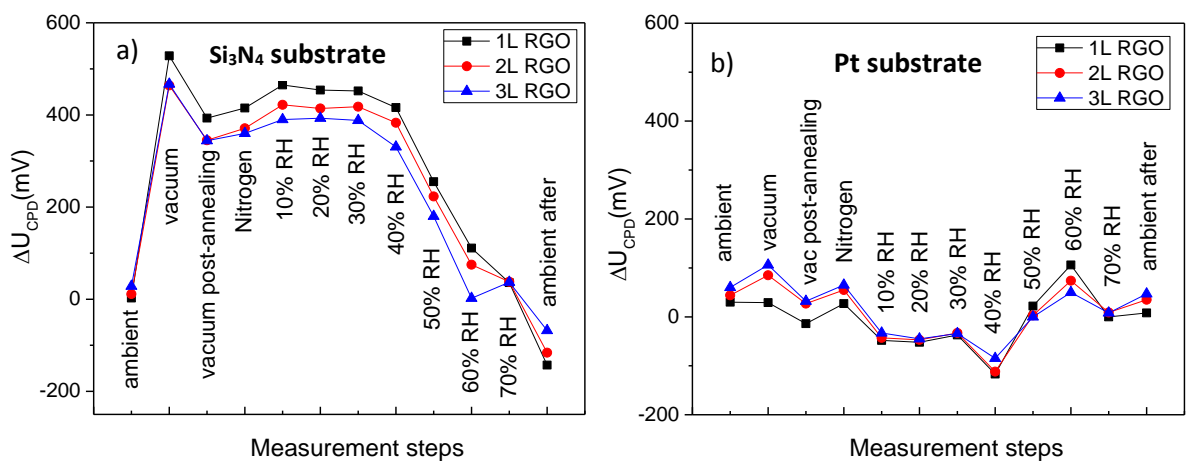

Figure 3: Summary of the relative change of surface potential values of RGO caused by the change in the humidity with respect to a) $Si_3N_4$ and b) Pt substrates. Absolute values for the surface potential of RGO and the substrate, as well as the substrate are included in the S.I. section.

We first analyse the case of RGO on $Si_3N_4$, where the greatest change in the surface potential (~500 mV) occurs upon the transition between ambient and vacuum (P ≈ 1 × 10$^{-5}$ mbar) conditions, corresponding to desorption of loosely attached airborne adsorbates, in accordance with previously published results on various types of graphene [46, 47, 48, 49, 50]. This shows that RGO is sensitive to the water and other gases present in the ambient that alter the electronic properties of RGO. It should be noted that at this transition the contrast of the surface potential swaps for RGO on $Si_3N_4$, with 1LRGO showing the lowest/highest values in ambient conditions and vacuum, respectively. This observation can be related to the fact that desorption of dopants occurs with different rate from domains of different thickness.

After thermal annealing in vacuum, at 150°C, the contrast between domains of different thickness becomes more pronounced and the edges become more defined. At the same time, the surface potential drops slightly in comparison to the vacuum step, which could be associated with an increase in sample conductivity due to further reduction of functional groups. Introduction of water vapour into the chamber leads only to small variations of the surface potential values up to 40% RH. However, the $\Delta U_{CPD}$ values sharply decrease with humidity in the range 40-70 % R.H., which could be attributed to the formation of a continuous layer of water and corresponding changes of work function.
For the majority of the environmental conditions studied here (i.e. from vacuum to 60 % R.H.), the relative surface potential remains highest for 1LRGO, decreasing with the number of layers. It is noteworthy that, the surface potential remains homogeneous within each thickness domain (Figure 1 b-e), which is most likely due to the relatively good conductivity of RGO. The image obtained at 70 % R.H. (Figure 1f), where the difference between the different domains becomes less clear might be attributed to a thin layer of water covering the whole sample and thus screening the underlying RGO at these conditions. This assumption is also confirmed by the fact that the contrast between individual domains is virtually lost at 70 % R.H.

The $\Delta U_{CPD}$ reaches the value initially observed in ambient at 60-70 % R.H., i.e. at humidity levels higher than typical ambient values of 40% RH. This fact indicates the importance of other dopants typically present in the ambient air that affect the electronic properties of graphene [48, 49, 50].

Contrary to effects discussed above, in the case of RGO on Pt, $\Delta U_{CPD}$ values stay almost independent on the humidity conditions, oscillating between -100 and +100 mV, with no clear effect of number of layers observed (Figure 2b).

*3.2 Graphene oxide – humidity variations*

The same experiment was conducted on the GO flakes. The image in Figure 4a shows the topography of the studied GO area. Different areas of the substrate, the border between them and the number of GO layers are indicated in the figure. Images 4b-f show surface potential maps of the GO area undergoing the same sequence of environmental conditions as shown previously for RGO.

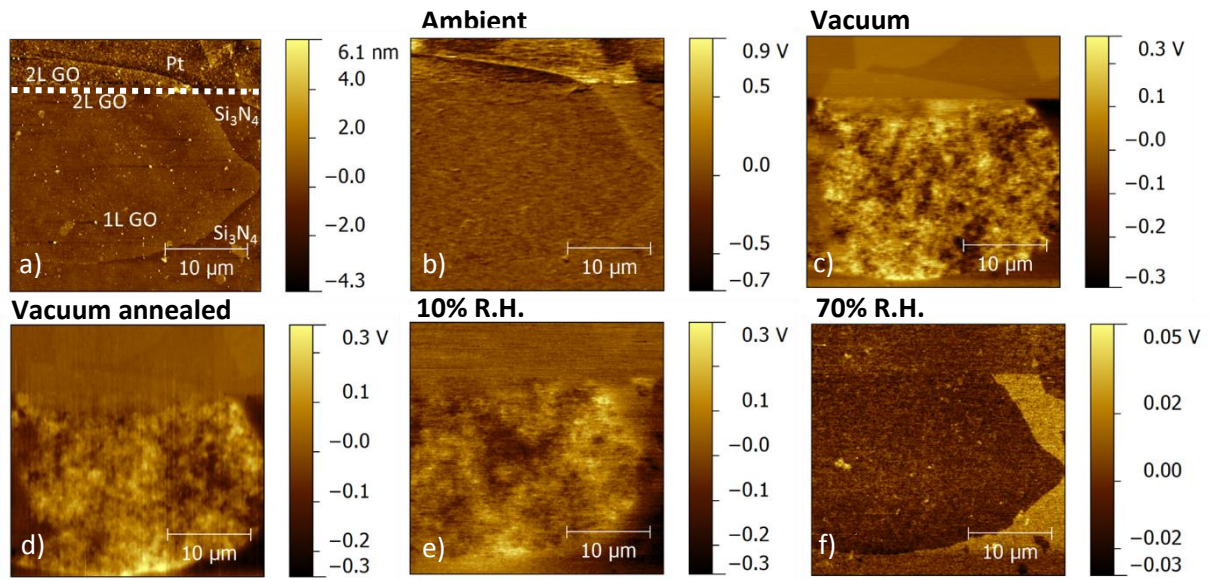

Figure 4: a) Topography and b-f) surface potential maps of GO on $Si_3N_4$ and Pt at different humidity levels: b) ambient; c) vacuum; d) post-thermal annealing in vacuum; e) 10% R.H. and f) 70% R.H. Underlying substrate and number of GO layers are shown in a). Dashed line in a) indicates border between $Si_3N_4$ and Pt areas of the supporting substrate. Note that the scale bars in b-f) are adapted to higlight individual features of the surface potential maps. All images are taken at room temperature.

The contact potential difference data extracted from surface potential maps are summarised in Figure 5 for both Si$_3$N$_4$ (Figure 5a) and Pt (Figure 5b) areas of the supporting substrate.

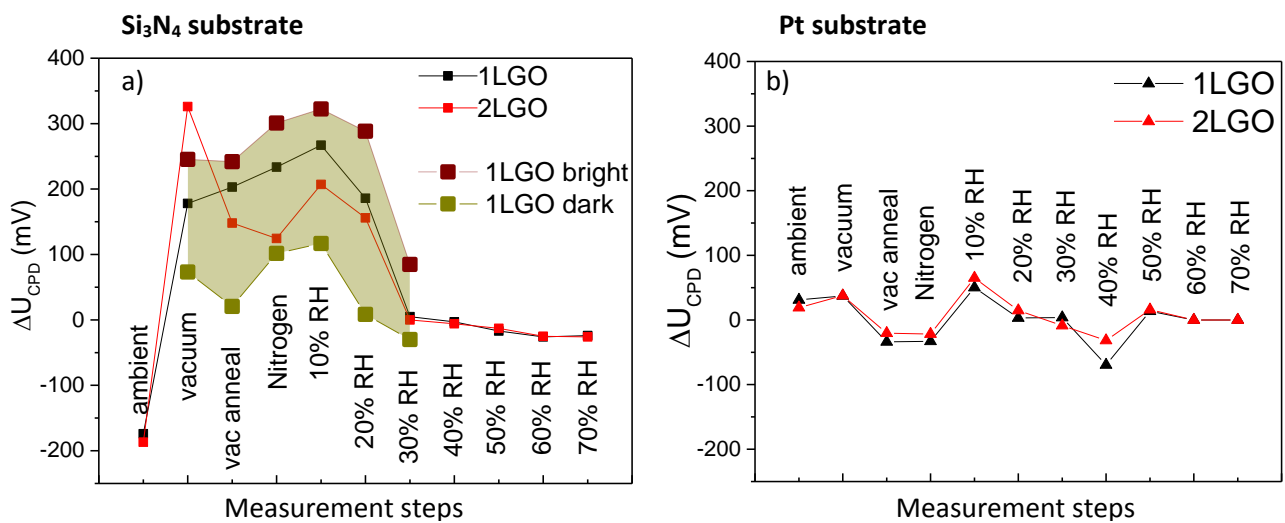

Figure 5: a) Summary of the relative change of the surface potential values of GO caused by the change in the humidity with respect to Si$_3$N$_4$. The surface potential values are average values over the 1 and 2LGO layer areas displayed in the corresponding image. Grey filled area shows the range of surface potential non-uniformities with respect to Si$_3$N$_4$. The associated values correspond to maximum brightness/darkness in each image. b) Summary of the relative change of the surface potential values of GO caused by the change in the humidity with respect to Pt. Absolute values for the surface potential of RGO and the substrate, as well as the substrate are included in the S.I. section.

In the case of GO flakes on Si$_3$N$_4$ (Figure 5a), the biggest change in $\Delta U_{CPD}$ occurs upon transition from ambient to vacuum (P $\approx$ 1 $\times$ 10$^{-5}$ mbar) conditions, similar to the behaviour observed for RGO, showing also the inversion of contrast between substrates and GO layers. This indicates that GO's electronic properties are modified by water, gases and other volatile organic contaminants present in the ambient air. The change from ambient to vacuum is more pronounced for 2LGO than for 1LGO. With the exception of the vacuum annealed and nitrogen steps, the overall trend observed with the change in humidity for 2LGO is similar with the one for 1LGO.

Contrary to the RGO case (Figure 2c), removal of atmospheric contaminants in vacuum gives rise to non-uniformities in the surface potential of the GO on Si$_3$N$_4$ scanned area and shows irregular domains of bright and dark contrast (Figure 4c), which are also observed at low humidity levels (Figure 4e). This is highlighted in Figure 5a, where the extreme values of the bright/dark patches are included for each measurement step from vacuum to 30% R.H. The bright and dark domains are attributed to charges trapped between the insulating GO and $Si_3N_4$ substrate. This is consistent with the removal of p-type adsorbates in vacuum, as well as removal of other functional groups, partly reducing the GO and making it electrostatically transparent to substrate charges as previously observed on RGO samples in ambient [38]. Consistent with screening effects associated with thicker layers, patches in 2LGO appear less intense compared to 1LGO in vacuum, become very faint in the vacuum annealed case (as impurities in the substrate are desorbed) and are completely absent for 70% RH due to screening effects associated with the water adsorbed on the GO sample.

The above claim is reaffirmed by the absence of this domain structure on GO/Pt area due to the metallic substrate acting as a drain electrode for charges. Above the 30% R.H. level, $\Delta U_{CPD}$ for RGO on Pt becomes again uniform showing almost constant values for all levels of humidity up to 70% R.H. Considering that GO is hydrophilic (i.e. more hydrophilic than RGO), it suggests that the GO flakes become completely covered by a water film at ~ 30% R.H., preventing additional changes at higher humidity levels. Indeed, as previously indicated, water forms hydrogen bonds with the hydroxyl and epoxy groups of GO, therefore absorbing significantly more water than pure graphene or RGO [51,52]. For the GO/Pt, $\Delta U_{CPD}$ values oscillate between $\pm$ 50 mV (Figure 5b) with no clear dependence on the number of layers, similar to RGO case, where however a larger spread of data ~$\pm$100 mV was measured (Figure 3b).

Summarising the two humidity experiments for GO and RGO supported on $Si_3N_4$, the main difference is that the surface potential reaches the values observed in ambient at different humidity levels: ~60% R.H for RGO and ~ 30% R.H. for GO, in line with the mainstream understanding that GO is more hydrophilic than RGO and formation of a continuous water layer occurs at lower humidity level on GO as compared to RGO due to the presence of functional groups. When deposited on Pt, apart from slightly larger surface potential variations for RGO, no significant differences are noted for both GO and RGO with the change in humidity, indicating that the electronic properties of GO and RGO on Pt are not affected by water vapours.

Our results clearly indicate that the substrate plays an important role and that complex mechanisms are governing the interaction between water molecules and GO and RGO. The role of the substrate was highlighted previously for pristine graphene on $Si/SiO_2$ [53], where it was shown that the effect of water strongly depends on the properties of the substrate. Here, the dipole moments of water adsorbates cause local electrostatic fields that can shift the substrate's defect states with respect to the Fermi energy of graphene and cause doping. Moreover, according to previous studies, it was postulated that the contact angle of water on graphene is dependent on both the liquid - graphene and liquid - substrate interactions, resulting in different degrees of wetting transparency of graphene [54]. As such, graphene is more transparent to wetting on hydrophilic substrates, but more opaque to wetting on hydrophobic substrates. It was also shown that few layer graphene does not significantly change the wettability of gold, silicon and copper, and few layer graphene appears hydrophilic in contradiction to bulk graphite [55]. The contact angles for Si and Au were measured ~ 32.6° and 77.4°, respectively, with a monolayer graphene coating causing only a ~1-2% increase in the contact angles when deposited on Si and Au [55]. $Si_3N_4$ versus Pt shows similar behaviour to Si versus Au, as the contact angles indicate hydrophilic character for $Si_3N_4$, with contact angle ~33° (similar to Si) [56] and 50° for Pt [57]. Following the same argument, it can thus be assumed that graphene, and also RGO, which has a reasonable structural and electrical resemblance to graphene, will show a more hydrophilic character on the $Si_3N_4$ substrate, which is more hydrophilic than Pt. This is in line with our observations of more pronounced changes in humidity for RGO placed on $Si_3N_4$ compared to Pt substrates. Although a similar trend is observed for GO, the analysis is more complicated as the functional groups make the surface strongly hydrophilic hindering the effect of the substrate.

Irrespective of the substrate, both GO and RGO show a similar trend for 1, 2 and 3 layers when subject to humidity variations, indicating that, for example for GO-based humidity sensors thickness is not an important factor, at least for the thickness range investigated in this paper. This is in contrast to studies of humidity effects on epitaxial graphene on SiC and CVD-grown graphene, which pointed to increased sensitivity with decreasing number of layers [48, 49, 50], with single-layer graphene being the most sensitive to environment changes.

### 3.3 *"Soft" reduction of GO during vacuum annealing*

As shown above, uniform surface potential maps are obtained for both GO and RGO on Pt, whereas non-uniform distribution $\Delta U_{CPD}$ is observed for GO on $Si_3N_4$, due to trapped charges. This is not, however, the case for RGO/ $Si_3N_4$, owing to a more conductive nature of RGO. To further investigate the nature of the observed patches, we study the effect of different vacuum annealing temperatures on the surface potential of GO. Figure 6a shows topography image of a GO area with marked regions of different number of layers. After obtaining topography and initial surface potential maps at room temperature (Figure 6b), the sample was sequentially annealed in vacuum for 1 hour at each temperature: 50°C, 100°C and 150°C, then allowed to cool down to room temperature to record the surface potential maps. Surface potential maps of GO for annealing temperatures of up to 100°C (Figure 6c-d) show almost no difference with respect to room temperature (Figure 6b), however a decrease in surface potential values is observed for 1 and 2LGO with increase of temperature. The surface potential of 1LGO changes at 150°C, where the patches disappear and the surface potential becomes effectively uniform as observed for RGO, likely to be consistent with GO chemical reduction when approaching 150°C. However, this effect is only temporary as re-exposing the sample to ambient after the vacuum thermal treatment at 150°C brings back the non-uniformities in surface potential (see details in S.I. section), due to either restored environmental doping trapped between the GO flake and the substrate or accumulated charges in $Si_3N_4$. The variation of GO surface potential as a result of annealing is furthermore illustrated by representative cross-sectional profiles taken at room temperature, 50°C, 100°C and 150°C for GO on $Si_3N_4$ (Figure 6g) and Pt substrates (Figure 6h).

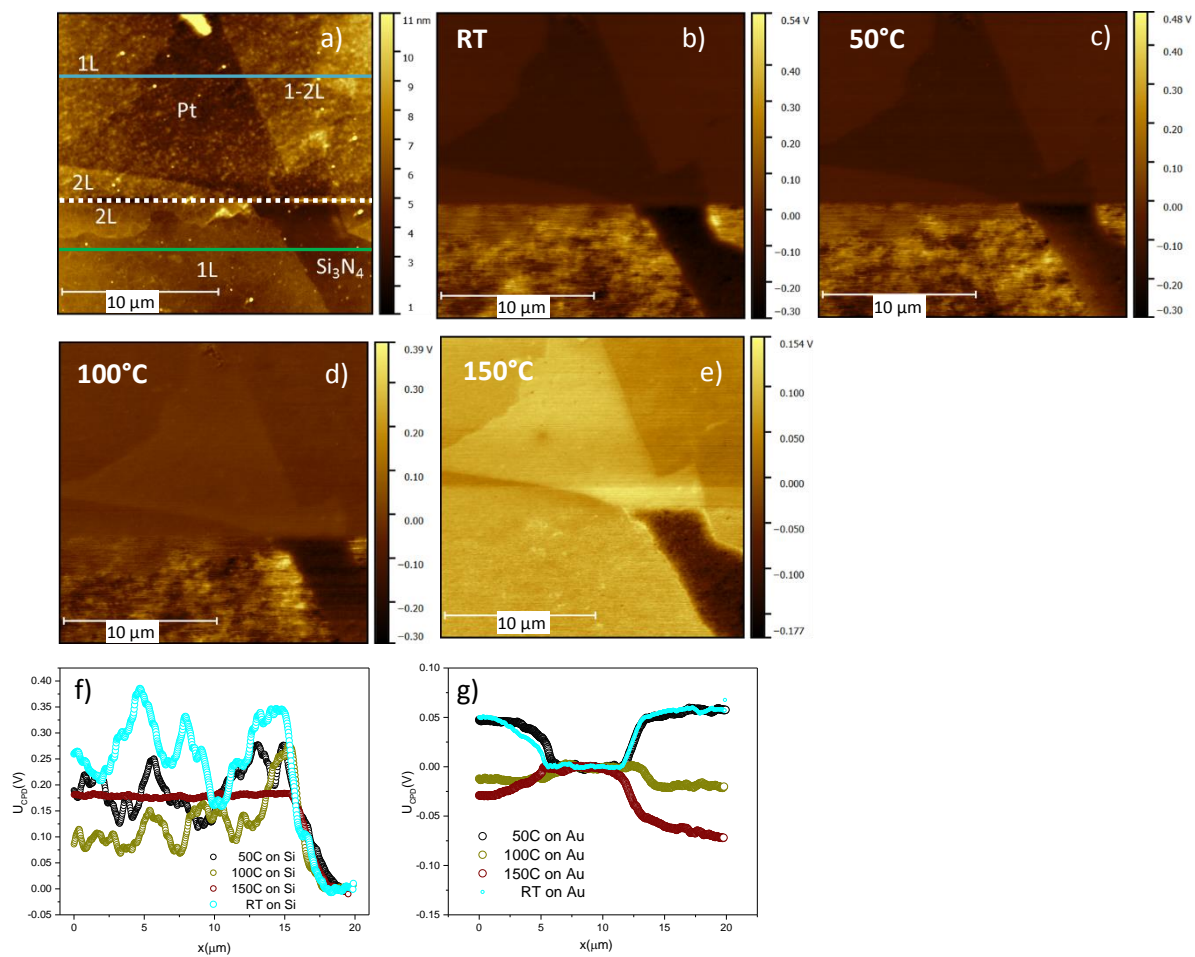

Figure 6: a) Topography; b-e) surface potential maps of GO taken in vacuum: b) initial state prior to annealing and c-e) after annealing at c) 50 °C, d) 100 °C, e) 150 °C. All measurements

were taken at room temperature. Dashed line in a) indicates a border between $Si_3N_4$ and Pt substrates. f), g) profiles of surface potential as measured at different temperatures. Profiles are taken along the solid blue lines (panel a): g) GO/ $Si_3N_4$ and h) GO/Pt.

**Conclusions**

The electronic properties of GO and RGO sheets deposited on metallic (Pt) and semiconducting ($Si_3N_4$) substrates have been investigated in controlled humidity environment. It is found that the supporting substrate plays an important role, with Pt making both GO and RGO insensitive to humidity variations and change in environment, from ambient to vacuum. Significant variations in surface potential with the humidity and environment are observed when GO and RGO sheets are placed on $Si_3N_4$, with the greatest change occurring for the transition between ambient and vacuum environment. The surface potential reaches the values observed initially in ambient at: ~60% R.H for RGO and ~ 30% for GO, in agreement with the conventional understanding that GO is more hydrophilic than RGO and that water saturation for GO occurs in lower humidity environment compared to that of RGO due to the presence of functional groups. Unlike epitaxial and CVD-grown graphene where the response to water and change in environment is strongly thickness dependent, GO and RGO do not exhibit significant thickness dependency for 1, 2 or 3 layers, indicating that the GO/RGO film thickness should not be a concern for humidity sensor applications. At the same time, we note that the electronic behaviour of GO and RGO is also affected by contaminants commonly present in ambient air, such as volatile organic compounds or gases and these effects should be taken into account for sensors designed to work in the ambient. The study provides valuable information for GO/RGO sensor design to ensure optimum sensitivity and reliability and will aid in developing models for realistic sensors.

**Acknowledgments** NPL authors acknowledge financial support from the UK's National Measurements Service under SC Graphene Project and the Graphene Flagship (no. CNECT-ICT-604391).

**Supplementary information**

1. SKPM on GO in ambient after thermal treatment in vacuum

SKPM in ambient was carried out on GO after thermal annealing in vacuum at 150 °C. The surface potential map and the topography is shown in Figure S1, where patches are observed in the surface potential map on the 1LGO region, indicating restoration of GO properties and only temporary effect of reduction.

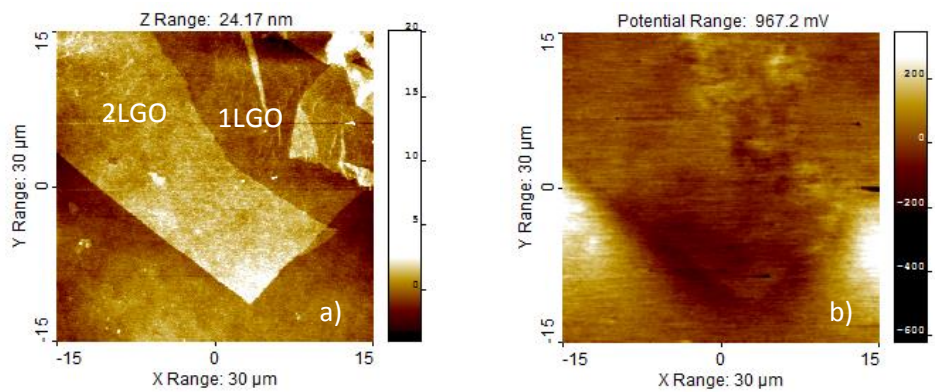

Figure S1: a) Topography and b) surface potential of GO in ambient after annealing of the sample at 150 °C.

2. Raman spectroscopy of GO and RGO (reduced at 300°C)

Raman intensity maps for the 2D-, G- and D-band were acquired on GO and RGO flakes supported on $Si_3N_4$ and are presented in Figure S2a and b.

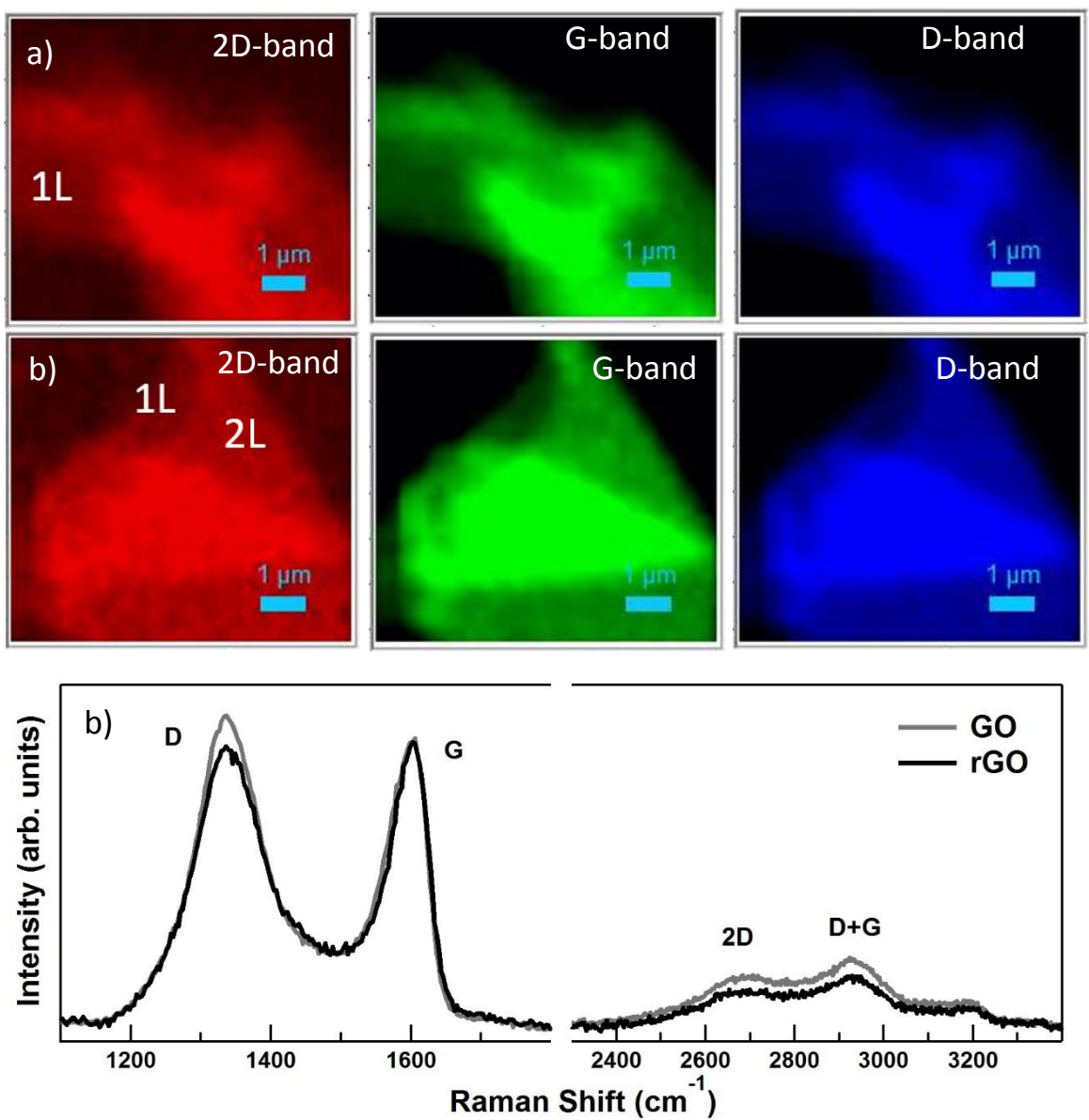

Figure S2: Raman intensity maps for a) GO and b) RGO flakes on $Si_3N_4$ substrate. The red intensity map is associated with the 2D peak, the green with the G peak and the blue with the D peak. (c) Representative Raman spectra for GO and RGO samples, showing D- and G-band region, and the 2D region.

All intensity maps display a uniform response for GO and RGO monolayer regions and show a clear increase in intensity with the layer thickness.

Individual Raman spectra extracted from preselected regions of monolayer flakes display two prominent peaks corresponding to the D- and G-band, respectively, as illustrated in Figure S2c. The position and full-width-at-half-maximum (FWHM) of the D- and G-peak does not depend on the number of layers. For GO, the D and G peaks appear at ~ 1341 cm$^{-1}$ and ~1593 cm$^{-1}$, and for RGO they appear at ~1342 cm$^{-1}$ and at ~1597 cm$^{-1}$, respectively. The D-peak position remains relatively constant upon annealing, whereas the G-peak shifts to higher frequency for RGO, which is consistent with charge transfer between the flake and the substrate. The increase in crystalline order upon annealing is supported by the decrease of the FWHM of the G, from 75 cm$^{-1}$ in the case of GO to 67cm$^{-1}$ for RGO. The intensity ratio of D- with respect to G-band is a measure of disorder and a slight decrease of $I_D/I_G$ is observed upon thermal reduction, from 1.1 to 0.9.

The 2D-band region for both GO and RGO shows three broad and low intensity peaks, which is significantly different from that of exfoliated monolayer graphene, dominated by a single and narrow 2D peak. Apart from the 2D peak, at ~ 2702 cm$^{-1}$ for GO, two additional peaks at ~ 2936 cm$^{-1}$ and at ~ 3202 cm$^{-1}$ are apparent, previously associated with disorder [38, 46, 47]. Upon reduction, peaks in the 2D region shift to lower frequency: ~2672 cm$^{-1}$ for the 2D peak and 2931 cm$^{-1}$ and 3189 cm$^{-1}$, respectively for the two additional contributions.

$U_{CPD}$ values for RGO on Pt and Si$_3$N$_4$ are summarised in Figure S3 and for GO in Figure S4.

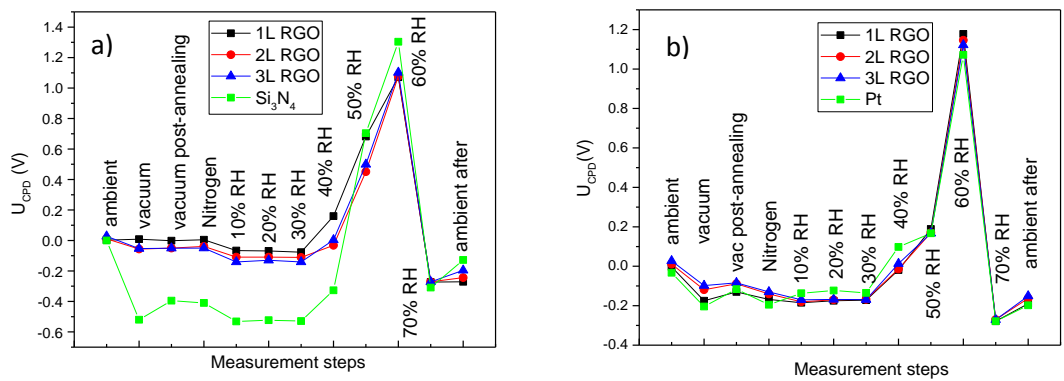

Figure S3 Absolute values for the measurement sequence of RGO on (a) Si$_3$N$_4$ and (b) Pt substrates.

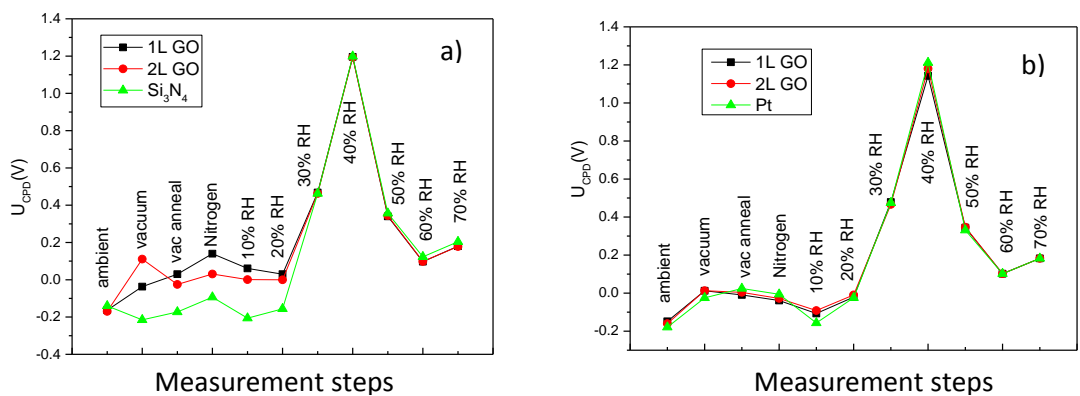

Figure S4 Absolute values for the measurement sequence of GO on (a) Si$_3$N$_4$ and (b) Pt substrates.